\documentclass[twocolumn]{aastex631}

\usepackage{amsmath}
\usepackage{array}
\usepackage{tabularx}
\usepackage{graphicx}

\begin{document}

\title{ALMA-QUARKS view of W49N: Multipolar episodic outflow associated with the most energetic Galactic water maser}

\author[0009-0005-9867-6723]{Yunfan Jiao}\thanks{E-mail: yunfan.astro@gmail.com}
\affiliation{Shanghai Astronomical Observatory, Chinese Academy of Sciences, No.80 Nandan Road, Shanghai 200030, People's Republic of China}
\affiliation{School of Astronomy and Space Sciences, University of Chinese Academy of Sciences,\\
No.19A Yuquan Road, Beijing 100049, People's Republic of China}

\author[0001-0002-5286-2564]{Tie Liu}\thanks{E-mail: liutie@shao.ac.cn}
\affiliation{State Key Laboratory of Radio Astronomy and Technology, Shanghai Astronomical Observatory, Chinese Academy of Sciences, \\
80 Nandan Road, Shanghai 200030, People's Republic of China}

\author[0000-0001-9822-7817]{Wenyu Jiao}
\affiliation{Shanghai Astronomical Observatory, Chinese Academy of Sciences, No.80 Nandan Road, Shanghai 200030, People's Republic of China}

\author[0000-0001-5950-1932]{Fengwei Xu}
\affiliation{Max Planck Institute for Astronomy, Königstuhl 17, 69117 Heidelberg, Germany}

\author[0000-0002-2826-1902]{Qilao Gu}
\affiliation{Shanghai Astronomical Observatory, Chinese Academy of Sciences, No.80 Nandan Road, Shanghai 200030, People's Republic of China}

\author[0000-0002-4154-4309]{Xindi Tang}
\affiliation{Xinjiang Astronomical Observatory, Chinese Academy of Sciences, Urumqi 830011, People's Republic of China}

\author[0000-0001-7573-0145]{Xiaofeng Mai}
\affiliation{Shanghai Astronomical Observatory, Chinese Academy of Sciences, No.80 Nandan Road, Shanghai 200030, People's Republic of China}

\author[0000-0003-4506-3171]{Qiuyi Luo}
\affiliation{Institute of Astronomy, Graduate School of Science, The University of Tokyo, 2-21-1 Osawa, Mitaka, Tokyo 181-0015, Japan}
\affiliation{Department of Astronomy, School of Science, The University of Tokyo, 7-3-1 Hongo, Bunkyo, Tokyo 113-0033, Japan}

\author[0000-0002-9836-0279]{Siju Zhang}
\affiliation{Departamento de Astronom\'{i}a, Universidad de Chile, Las Condes, 7591245 Santiago, Chile}

\author[0000-0002-6622-8396]{Paul F. Goldsmith}
\affiliation{Jet Propulsion Laboratory, California Institute of Technology, 4800 Oak Grove Drive, Pasadena, CA 91109, USA}

\author[0000-0002-3179-6334]{Chang Won Lee}
\affiliation{Korea Astronomy and Space Science Institute (KASI), 776 Daedeokdae-ro, Yuseong-gu, Daejeon 34055, Republic of Korea}
\affiliation{University of Science and Technology, Korea (UST), 217 Gajeong-ro, Yuseong-gu, Daejeon 34113, Republic of Korea}

\author[0000-0003-1649-7958]{Guido Garay}
\affiliation{Departamento de Astronom\'{i}a, Universidad de Chile, Las Condes, 7591245 Santiago, Chile}
\affiliation{Chinese Academy of Sciences South America Center for Astronomy, National Astronomical Observatories, CAS, Beijing 100101, China}

\author[0009-0000-0178-7472]{Yuhan Yang}
\affiliation{Shanghai Astronomical Observatory, Chinese Academy of Sciences, No.80 Nandan Road, Shanghai 200030, People's Republic of China}
\affiliation{School of Astronomy and Space Sciences, University of Chinese Academy of Sciences,\\
No.19A Yuquan Road, Beijing 100049, People's Republic of China}

\author[0000-0003-1602-6849]{Prasanta Gorai}
\affiliation{Rosseland Centre for Solar Physics, University of Oslo, PO Box 1029 Blindern, 0315 Oslo, Norway}
\affiliation{Institute of Theoretical Astrophysics, University of Oslo, PO Box 1029 Blindern, 0315 Oslo, Norway}

\author[0000-0003-0709-708X]{Manuel Merello}
\affiliation{Departamento de Astronom\'{i}a, Universidad de Chile, Casilla 36-D, Santiago, Chile}
\affiliation{Centro de Astro-Ingeniería (AIUC), Pontificia Universidad Católica de Chile, Av. Vicuña Mackena 4860, Macul, Santiago, Chile}

\author[0000-0002-8586-6721]{Pablo García}
\affiliation{Chinese Academy of Sciences South America Center for Astronomy, National Astronomical Observatories, CAS, Beijing 100101, China}
\affiliation{Instituto de Astronom\'ia, Universidad Cat\'olica del Norte, Av. Angamos 0610, Antofagasta, Chile}

\author[0000-0002-8697-9808]{Sami Dib}
\affiliation{Max Planck Institute for Astronomy, Königstuhl 17, 69117 Heidelberg, Germany}

\author[0000-0001-7866-2686]{Jihye Hwang}
\affiliation{Institute for Advanced Study, Kyushu University, 774 Motooka nishi-ku, Fukuoka, Japan}
\affiliation{Department of Earth and Planetary Sciences, Faculty of Science, Kyushu University, nishi-ku, Fukuoka 819-0395, Japan}

\author[0009-0003-6633-525X]{Ariful Hoque}
\affiliation{S. N. Bose National Centre for Basic Sciences, Block-JD, Sector-III, Salt Lake City, Kolkata 700106, India}

\author[0000-0002-5809-4834]{Mika Juvela}
\affiliation{Department of Physics, P.O. box 64, FI- 00014, University of Helsinki, Finland}

\author[0000-0001-7817-1975]{Yankun Zhang}
\affiliation{Shanghai Astronomical Observatory, Chinese Academy of Sciences, No.80 Nandan Road, Shanghai 200030, People's Republic of China}

\author[0000-0002-7125-7685]{Patricio Sanhueza}
\affiliation{Department of Astronomy, School of Science, The University of Tokyo, 7-3-1 Hongo, Bunkyo, Tokyo 113-0033, Japan}

\author[0009-0008-8439-8488]{Jixiang Weng}
\affiliation{Shanghai Astronomical Observatory, Chinese Academy of Sciences, No.80 Nandan Road, Shanghai 200030, People's Republic of China}

\author[0000-0003-2412-7092]{Kee-Tae Kim}
\affiliation{Korea Astronomy and Space Science Institute (KASI), 776 Daedeokdae-ro, Yuseong-gu, Daejeon 34055, Republic of Korea}
\affiliation{University of Science and Technology, Korea (UST), 217 Gajeong-ro, Yuseong-gu, Daejeon 34113, Republic of Korea}

\author[0000-0002-9836-0279]{Swagat R. Das}
\affiliation{Departamento de Astronom\'{i}a, Universidad de Chile, Las Condes, 7591245 Santiago, Chile}

\author[0000-0002-6386-2906]{Archana Soam}
\affiliation{Indian Institute of Astrophysics, II Block, Koramangala, Bengaluru, 560034, India}

\author[0000-0003-0295-6586]{Tapas Baug}
\affiliation{S. N. Bose National Centre for Basic Sciences, Block-JD, Sector-III, Salt Lake City, Kolkata 700106, India}

\author[0000-0003-0356-818X]{Jianjun Zhou}
\affiliation{Xinjiang Astronomical Observatory, Chinese Academy of Sciences, Urumqi 830011, People's Republic of China}

\author[0000-0002-9574-8454]{Leonardo Bronfman}
\affiliation{Departamento de Astronom\'{i}a, Universidad de Chile, Las Condes, 7591245 Santiago, Chile}

\author[0000-0003-4546-2623]{Aiyuan Yang}
\affiliation{National Astronomical Observatories, Chinese Academy of Sciences, Beijing 100101, People's Republic of China}
\affiliation{Key Laboratory of Radio Astronomy and Technology, Chinese Academy of Sciences, A20 Datun Road, Chaoyang District, Beijing, 100101, People's Republic of China}

\author{Lei Zhu}
\affiliation{Chinese Academy of Sciences South America Center for Astronomy, National Astronomical Observatories, CAS, Beijing 100101, China}

\begin{abstract}
We present a detailed investigation of a multipolar episodic molecular outflow in the mini-starburst region W49N, which hosts the most luminous water maser in the Galaxy. Using high-resolution ($\sim0\farcs3$) Atacama Large Millimeter/submillimeter Array (ALMA) observations of the $\mathrm{^{12}CO}$ emission as part of the ALMA-QUARKS survey, we analyze the morphology and kinematics of the outflow. Our observations reveal four newly identified outflow lobes in addition to the previously known central bipolar jet. These lobes appear more jet-like rather than exhibiting wide opening angles. Based on the $\mathrm{^{12}CO}$ (2--1) and $\mathrm{^{13}CO}$ (2--1) emission, we provide a more reliable estimate of the outflow's physical parameters, confirming it as one of the most energetic outflows in the Galaxy. Notably, these newly discovered lobes exhibit chains of knots, a characteristic signature of episodic ejection. Furthermore, two of the lobes display prominent S-shaped wiggles, suggestive of a precessing jet. The discovery of these features---commonly observed in outflows from low-mass protostars---in such an extreme massive star-forming environment provides compelling evidence that some underlying physical mechanisms for launching outflows are conserved across a wide range of stellar masses.
\end{abstract}

\keywords{stars: formation---stars: protostars---ISM: jets and outflows}

\section{Introduction} \label{sec:intro}
By carrying angular momentum away from infalling material, protostellar outflows serve as large-scale observational probes of the deeply embedded accretion and ejection mechanisms, preserving a dynamical archive of the star formation history \citep{2016Bally,2020A&ARv..28....1L}. Discontinuous or knotty structures in these outflows are common manifestations of time-variable ejections, a phenomenon first identified in early studies of Herbig-Haro objects \citep[e.g.,][]{1989Reipurth}. While such morphological discontinuities can sometimes arise from interactions with a clumpy ambient medium \cite[e.g.,][]{1999deGouveia,2002Raga} or intrinsic magnetic field and launching instabilities \cite[e.g.,][]{2023Shang}, highly regular time variability is widely thought to originate from episodic accretion onto the protostar, which in turn induces variations in outflow velocity and mass-loss rate \citep{2007Arce}. Consequently, faster material ejected at a later time catches up with the slower preceding ejecta, forming an internal shock \citep[see the review by][]{2020A&ARv..28....1L}. This internal interaction gives rise to the characteristic observational signature of an episodic outflow: a chain of luminous, regularly spaced knots along its trajectory \citep{1990Raga,2004Lee,2022Jhan}.

Knotty structures are commonly observed within bipolar outflows from low-mass protostars in relatively isolated environments \cite[for example, see][]{2010ApJ...717...58H,2016ApJ...816...32J,2017NatAs...1E.152L,2020A&ARv..28....1L,2024AJ....167...72D,2025ApJ...979...17L}, a subset of which display S-shaped trajectories suggestive of precessing jets \citep{2007Lee,2008Rodriguez,2016Podio}. In recent years, episodic jets have also been discovered in cluster environments \citep[e.g., CARMA-7 in][]{2015Plunkett} and in massive star-forming regions (e.g., W43-MM1 in \citealt{2020Nony} and SDC335 in \citealt{xu2023, zou2025}). As estimated in \cite{2020Nony}, the period of the episodic outflow driven by the massive protostar in W43 is about 500 years, which is similar to the timescales measured for low-mass protostars. However, due to observational constraints, particularly angular resolution, such objects remain scarce, and the underlying mechanisms driving periodic jets in massive stars remain unclear. 

To address this knowledge gap, it is essential to study molecular outflows driven by massive protostars in various environments. The mini-starburst region W49N (IRAS 19078+0901), located at a distance of $\sim$ 11.11 kpc from the Sun \citep{2013ApJ...775...79Z}, presents an exceptional opportunity. This region hosts a rich population of young massive stars and ultracompact {H\textsc{ii}} regions, including at least two O-type protostars within source G of W49N with estimated masses of $\sim$ 20--30 $\mathrm{M_{\odot}}$ \citep{2020DePree}. Furthermore, W49N-G drives the most luminous water maser in the Galaxy, a robust signpost of intense molecular outflow activity that is responsible for the masers' excitation \citep{1992Gwinn}. The water masers exhibit extremely powerful flare phenomena \citep{2019A&A...628A..89V,2020MNRAS.496L.147V,2023MNRAS.522L...6V}, which are rarely seen in other star-forming regions and may be in connection with a massive close binary star system \citep{2020MNRAS.496L.147V}. These observations make W49N a unique target for investigating molecular outflows in extreme environments. However, the molecular outflows driven by source G have remained relatively underexplored. Initially, \cite{1986Scoville} mapped the CO (1--0) emission using the Owens Valley Radio Observatory (OVRO) millimeter-wave interferometer and the Five College Radio Astronomy Observatory (FCRAO) 14-m telescope under a resolution of $\sim7\arcsec$. They identified a powerful bipolar outflow likely driven by ionized stellar winds. Building on this, work by \cite{2015LiuTie} utilized Submillimeter Array (SMA) observations of the HCN (3--2) line to reveal a compact bipolar outflow, identifying it as one of the most energetic outflows in the Milky Way, while still limited in resolution ($>2\arcsec$ or $>0.1$ pc). In addition, the physical parameters of the outflow calculated using HCN emission are highly uncertain because HCN abundance in outflows could be significantly enhanced by shocks \citep{2020A&A...634A..17J}.

In this letter, we utilize the unprecedented angular resolution and sensitivity of the Atacama Large Millimeter/submillimeter Array (ALMA) to conduct a detailed investigation of the molecular outflows in W49N. By analyzing the $\mathrm{^{12}CO}$ (2--1) and $\mathrm{^{13}CO}$ (2--1) emission, which have more stable relative abundances and can probe the outflow more comprehensively, we aim to place better constraints on the morphology, kinematics, and physical properties of the outflows, and to further shed new light on the ejection processes associated with the most massive young protostars.

\section{Observations} \label{sec2}
W49N is one of the 139 massive clumps observed in the ALMA project ``Querying Underlying mechanisms of massive star formation
with ALMA-Resolved gas Kinematics and Structures" (QUARKS; PIs: Lei Zhu, Guido Garay, and Tie Liu). Band 6 observations were conducted using the ALMA 7-m array (the Atacama Compact Array, or ACA), as well as the 12-m array with configurations C-2 and C-5 (ALMA Project Code: 2021.1.00095.S; PI: Lei Zhu). Four spectral windows (SPWs) were set with bandwidths of 1.875 GHz, in which SPW2 (centered at 220.319 GHz) and SPW3 (centered at 231.370 GHz) cover molecular outflow tracers $\mathrm{^{13}CO}$ (2--1) (rest frequency: 220.399 GHz) and $\mathrm{^{12}CO}$ (2--1) (rest frequency: 230.538 GHz), respectively. To derive the outflow parameters and balance the detection sensitivity and angular resolution, we smoothed the two molecular-line data cubes to a common beam size of $0\farcs53\times0\farcs43$. The corresponding spatial resolution for W49N is $\sim 5300\ \mathrm{AU}$. The noise level is $\sim 5\ \mathrm{mJy\ beam^{-1}}$ per 976.563 kHz ($\sim 1.27\ \mathrm{km\ s^{-1}}$) channel. Details on the observation setup and data reduction of the QUARKS survey are described in \cite{2024Liu, xu2024}.

We also retrieved additional high-resolution archival 1.3 $\mathrm{mm}$ continuum data (Project Code: 2018.1.00520.S; PI: David Wilner) from the ALMA Science Archive, observed from 2019 July 11 to 2019 July 13 at Band 6. The data combined four SPWs covering 240.8--244.5 and 255.9--259.6 $\mathrm{GHz}$, resulting in an effective central frequency of 250.16 GHz and a total bandwidth of 7.5 $\mathrm{GHz}$. The synthesized beam size is $\mathrm{0.038\arcsec\times 0.032\arcsec}$ ($\mathrm{PA=79.5\arcdeg}$), corresponding to a physical resolution of $\sim$ 300 AU at the distance of W49N, which were used to identify the outflow driving source. The noise level for continuum emission is $\sim 0.35\ \mathrm{mJy\ beam^{-1}}$.

The 22.2 GHz water maser data were drawn from the first-epoch observations of the Bar and Spiral Structure Legacy (BeSSeL) Survey \citep{2011Brunthaler}, a Key Science Project of the National Radio Astronomy Observatory \footnote{The National Radio Astronomy Observatory is a facility of the National Science Foundation operated under cooperative agreement by Associated Universities, Inc.}. The observation was conducted with the Very Long Baseline Array (VLBA) on 2010-3-13. The channel width is 31.25 kHz, corresponding to a velocity resolution of $\sim 0.42\ \mathrm{km\ s^{-1}}$. We selected maser spots brighter than 30 $\mathrm{Jy\ beam^{-1}}$, which is $\sim0.3\%$ of the flux density of the brightest spot. The BeSSeL Survey data and objectives are described on their project website \footnote{\url{https://bessel.vlbi-astrometry.org}}, and detailed observation setups and calibration procedures for masers can be found in \cite{2013ApJ...775...79Z}. 

\section{Results} \label{sec3}

\subsection{Morphology of Jets and Outflows}
As an ideal molecular outflow tracer, CO can distinctly reveal the directions, velocities, knotty structures, and cavities of protostellar outflows \citep[for example, ][]{1996Bachiller,2007Arce,2016Bally}. In this section, we conduct a thorough investigation of the morphology of the molecular outflows in W49N using $\mathrm{^{12}CO}$ (2--1) and $\mathrm{^{13}CO}$ (2--1) emission lines. Hereafter, we abbreviate these two lines as $\mathrm{^{12}CO}$ and $\mathrm{^{13}CO}$, respectively.

\begin{figure*}[ht!]
\centering
\includegraphics[width=1\linewidth]{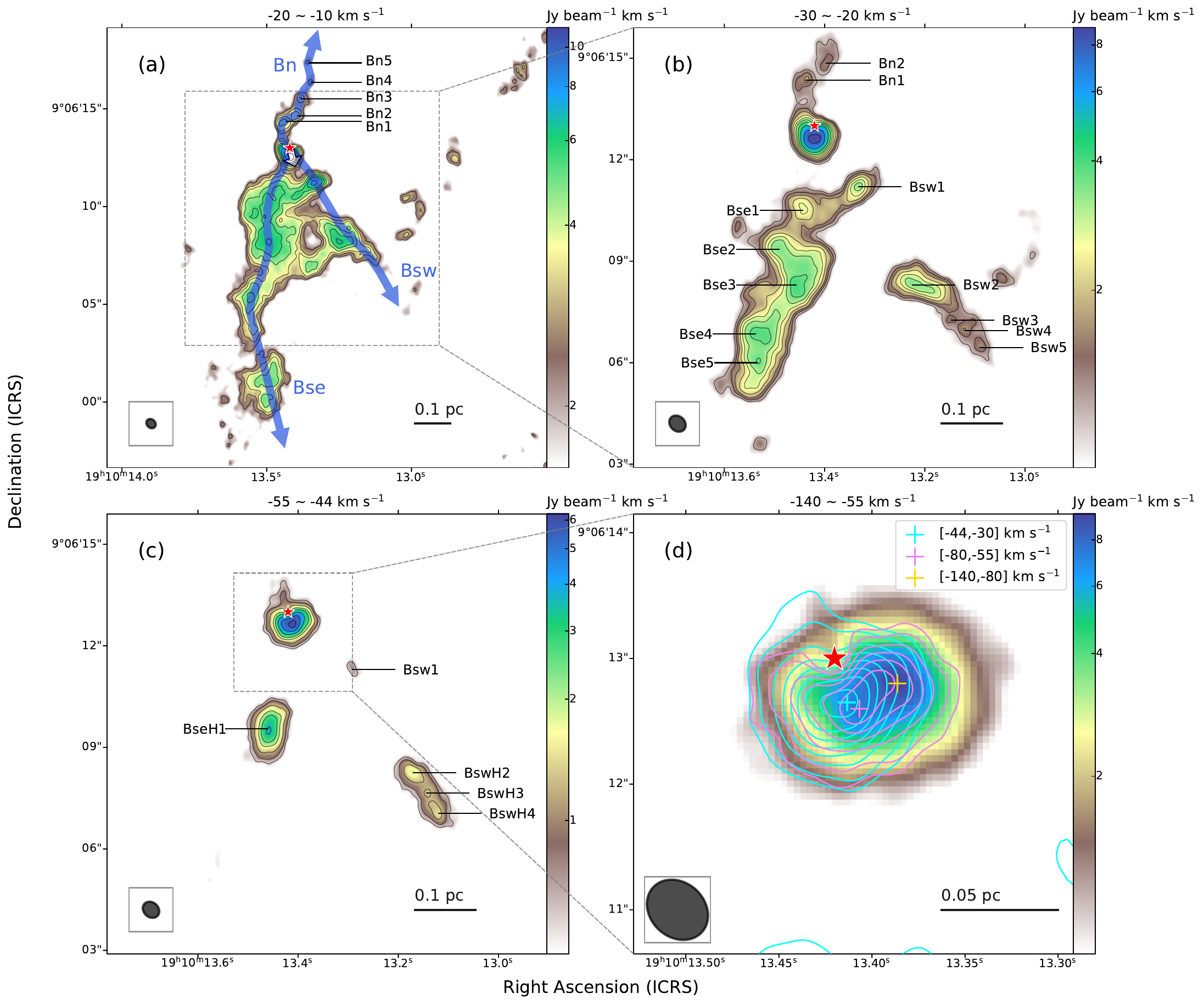}
\caption{\em{Moment 0 maps of blueshifted $^{12}CO$ emission in four velocity channels, shown in both color scales and superimposed contours. Panel \textbf{(a), (b), (c)} and \textbf{(d)} correspond to low-, medium-, high- and extremely high-velocity channels. The levels of black contours are [7, 9, 11, 14, 17, 20, 24, 30, 42, 55] times the rms noise for panel \textbf{(a)} and \textbf{(b)}, and [7, 10, 15.5, 24, 32, 45, 60, 75] for panel \textbf{(c)}, where the rms in each panel is estimated from background regions in the corresponding channel. Beams and the scale bars are shown in the bottom left and right corners, respectively. In panel \textbf{(a), (b)} and \textbf{(c)}, distinct knots are labeled following "B" for blueshifted + direction + "H" for high-velocity in panel \textbf{(c)} + number in order of distance from the central source for a given direction. The red star marks the location of the central protostar, determined from the peak of the continuum emission. Four outflow lobes are illustrated by four blue arrows in panel \textbf{(a)}. The short white arrow with a dark blue edge represents the previously studied outflow lobe nearly along the line of sight \citep{2015LiuTie}. Panel \textbf{(d)} shows contours and emission peaks (plus signs) of the inner region from diverse channels. The extremely high-velocity component is divided into [-80, -56] and [-140, -80] $\mathrm{km\ s^{-1}}$ to show the velocity-dependent spatial distribution. Note that the emission peak tends to move outward as velocity increases.
}}
\label{fig:channel_b}
\end{figure*}

\begin{figure*}[ht!]
\centering
\includegraphics[width=1\linewidth]{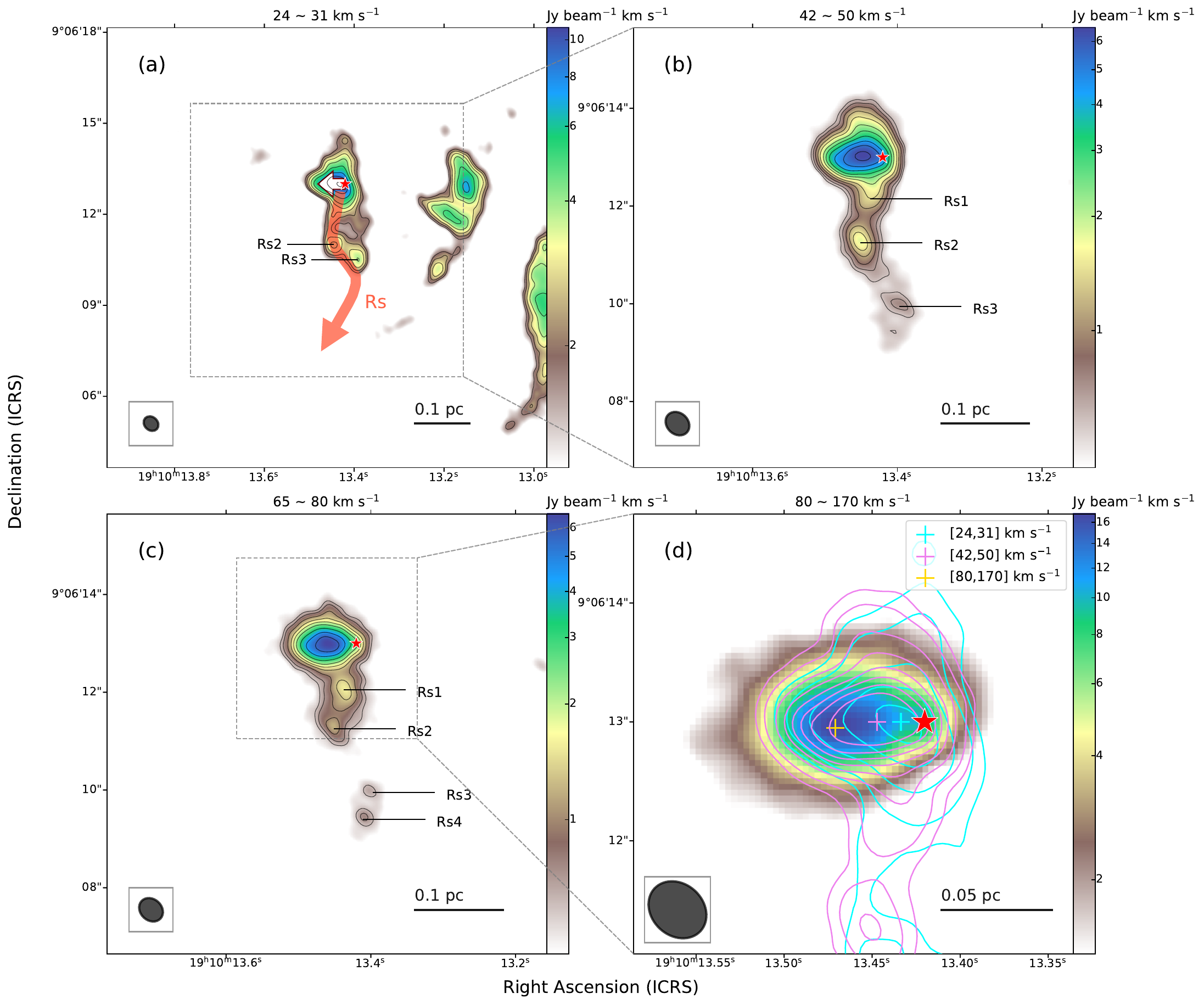}
\caption{\em{
Similar to \autoref{fig:channel_b}, here ``R" denotes redshifted part.
}}
\label{fig:channel_r}
\end{figure*}

We adopt a systemic velocity of $v_{sys}=+6.2\ \mathrm{km\ s^{-1}}$ for W49N \citep{2024Liu}. To analyze the outflow structures, we divided the emission into low-, medium-, high-, and extremely high-velocity components by visually inspecting the $\mathrm{^{12}CO}$ spectrum. For the redshifted wing, we selected the line-of-sight (LOS) channel velocity ($v_{LOS}$) intervals [24, 31], [42, 50], [65, 80], and [80, 170] $\mathrm{km\ s^{-1}}$ specifically to mitigate absorption effects (in [31, 42] and [50, 65] $\mathrm{km\ s^{-1}}$). The redshifted absorption may arise in diffuse foreground clouds along the sight-line to W49N, which has also been detected in other molecular lines against the bright continuum sources \citep{2015Liszt,2015Neufeld,2024Luo}. In contrast, the blueshifted wing is absorption-free. Instead of mirroring the redshifted intervals, we defined ranges of [-20, -10], [-30, -20], [-55, -44] and [-140, -55] $\mathrm{km\ s^{-1}}$ to better delineate the knotty structures. The moment 0 maps of these velocity channels are presented in \autoref{fig:channel_b} and \autoref{fig:channel_r}. The intermediate velocity range [-44, -30] $\mathrm{km\ s^{-1}}$ in the blueshifted wing represents a transition between different velocity components. Its compact part near the source is visualized as contours overlaid on the extremely high-velocity map to illustrate the positional shift of emission peaks.

In contrast to the previously unresolved extremely high-velocity jet traced by HCN (3--2) emission \citep{2015LiuTie}, the $\mathrm{^{12}CO}$ line emission has newly revealed three blueshifted lobes (``Bn", ``Bse", ``Bsw") and one redshifted lobe (``Rs"), predominantly aligned in the north--south direction. These are indicated by the curved arrows in panel (a) of \autoref{fig:channel_b} and \autoref{fig:channel_r}. As shown in panel (d) of both figures, the central bipolar jet is fully resolved in the $\mathrm{^{12}CO}$ data, exhibiting two compact blobs aligned roughly east--west (see also panel (c) in \autoref{fig:wiggle}). These blobs exhibit LOS flow velocities of up to 160 $\mathrm{km\ s^{-1}}$ relative to the systemic velocity, while the redshifted component likely reaches even higher speeds but extends beyond our SPW coverage. Moreover, they show clear outward motion with increasing flow velocities, suggesting a highly collimated jet inclined out of the plane of sky, whose intrinsic velocity is dominated by the LOS component. Its inclination angle between the flow axis and the plane of sky was estimated to be $53\arcdeg\pm13\arcdeg$ through analyzing water maser features \citep{2023Asanok}.

Panel (c) of \autoref{fig:channel_b} and \autoref{fig:channel_r} further reveal that the two blueshifted outflow lobes on the southern side, along with the newly identified redshifted lobe, reach flow velocities of $\sim$ -60 and $\sim$ 70 $\mathrm{km\ s^{-1}}$, respectively. Given their extended projected lengths, which suggest relatively small inclination angles to the plane of sky, the actual intrinsic velocities of these newly identified lobes could be roughly comparable to those of the central, compact jets.

It is worth noting that the four newly discovered outflow lobes are relatively collimated rather than wide-angle lobes. Furthermore, each outflow lobe exhibits a chain of knotty structures, as marked in \autoref{fig:channel_b} and \autoref{fig:channel_r}. In addition, prominent outflow wiggles can be seen in two outflow lobes ``Bn" and ``Rs" (see panels (a) and (b) in \autoref{fig:wiggle}). The details are discussed in \autoref{sec4}.

\subsection{Outflow parameters}
Given the multiple newly identified outflow lobes in different directions, it is essential to recalculate outflow mass, momentum, and energy to provide a more accurate understanding of their dynamics and physical properties. To reduce the impact of optical depths in low-velocity channels of $\mathrm{^{12}CO}$ emission, we apply two methods for calculating outflow parameters: (1) correcting the optical depth with both $\mathrm{^{12}CO}$ and $\mathrm{^{13}CO}$ emissions, and (2) using only $\mathrm{^{13}CO}$ which is more optically thin.

For optically thin linear molecules, the total column density follows
\begin{equation}
    N = \frac{3c^2}{16\pi^3\Omega_s S\mu^2\nu^3}\left(\frac{Q_{rot}}{g_J g_K g_I}\right)exp\left(\frac{E_u}{kT}\right)\int{S_\nu dv}
    \label{eq:column_density}
\end{equation} 
from \cite{2015Mangum} and the outflow mass follows
\begin{equation}
    M = \frac{N\Omega_s D^2}{X}\cdot \mu_{\mathrm{H_2}}m_{\mathrm{H}},
    \label{eq:gas_mass}
\end{equation}
where $c$ is the speed of light, $m_{\mathrm{H}}$ is the mass of a hydrogen atom, and $\Omega_s$, $S$, $\mu$, $\nu$, $E_u$, $D$, $X$, $\mu_{\mathrm{H_2}}$, $S_\nu$ denote solid angle, line strength, dipole moment, rest frequency, upper level energy, source distance, abundance ratio (CO-to-$\mathrm{H_2}$), mean molecular weight per hydrogen molecule and flux density, respectively. For diatomic linear molecules, rotational degeneracy is $g_J = 2J+1$, where $J$ is the lower level; $K$ and the spin degeneracies are $g_K = g_I = 1$. Under the LTE assumption where all levels are populated with a single excitation temperature T, the rotational partition function $Q_{rot}\approx kT/(hB_0)+1/3$, where $k$ and $h$ are Boltzmann's and Planck's constant, and $B_0$ is the rigid rotor rotation constant, with values of 57635.96 and 55101.01 $\mathrm{MHz}$ for $\mathrm{^{12}CO}$ and $\mathrm{^{13}CO}$, respectively. We adopt a standard [$\mathrm{^{12}CO}$]/[$\mathrm{H_2}$] relative abundance of $10^{-4}$ \citep{2013Bolatto}, and a mean molecular weight ($\mu_{\mathrm{H_2}}$) of 2.8 for a molecular cloud made with $N(\mathrm{H_2})/N(\mathrm{He})=10$ and negligible metals \citep{2008Kauffmann}.

\begin{table*}[ht!]
    \centering
    \caption{Outflow Parameters Derived from Three Different Methods}
    \label{tab:OutflowParameters}
    \begin{tabular*}{\textwidth}{@{\extracolsep{\fill}}C|C C C|C C C|C C C}
        \hline
        \hline
        \mathrm{v_{LOS}} & \mathrm{M_{12}} & \mathrm{M_{12\tau}} & \mathrm{M_{13}} & \mathrm{P_{12}} & \mathrm{P_{12\tau}} & \mathrm{P_{13}} & \mathrm{E_{12}} & \mathrm{E_{12\tau}} & \mathrm{E_{13}} \\
        
         \mathrm{(km\ s^{-1})} & &\mathrm{(M_{\odot})}  &  & &\mathrm{(M_{\odot}\ km\ s^{-1})}  &  & & \mathrm{(10^{47}\ erg)}  &   \\
        
        \hline
        [-140,-55] & 1.3 & - & - & 120 & - & - & 1.3 & - & - \\
        
        [-55,-20] & 7.3 & 15 & 17 & 280 & 520 & 600 & 1.1 & 2.0 & 2.2 \\

        [-20,-10] & 14 & 92 & 103 & 300 & 2000 & 2200 & 0.65 & 4.2 & 4.6 \\
        
        [24,31] & 2.6 & 10 & 13 & 54 & 210 & 260 & 0.11 & 0.43 & 0.51 \\
        
        [42,50] & 0.86 & 1.2 & 2.9 & 34 & 47 & 110 & 0.13 & 0.18 & 0.43 \\
        
        [65,170] & 2.0 & - & - & 200 & - & - & 2.1 & - & - \\
        
        \hline
    \end{tabular*}
\end{table*}

According to \cite{2009Qiu}, the derivation of an outflow's gas mass using $\mathrm{^{12}CO}$ emission is given by the equation
\begin{equation}
M_{12} = 1.39\times10^{-6}e^{\frac{16.59}{T}}(T+0.92)D^2\int{\frac{\tau_{12}}{1-e^{-\tau_{12}}}S_\nu dv},
\end{equation}
where $\tau_{12}$ represents the optical depth of $\mathrm{^{12}CO}$, and $M$, $D$, $S_\nu$ are in units of $\mathrm{M_\odot}$, $\mathrm{kpc}$ and $\mathrm{Jy}$, respectively. Assuming that the excitation temperatures of $J=$ 2--1 transition of the two isotopologues $\mathrm{^{12}CO}$ and $\mathrm{^{13}CO}$ are the same, the average optical depth can be calculated by their flux in each velocity channel following
\begin{equation}
    \frac{S_\nu(\mathrm{^{12}CO})}{S_\nu(\mathrm{^{13}CO})} = \frac{1-e^{-\tau_{12}}}{1-e^{-\tau_{12}/\chi}},
\end{equation}
where $\chi = (7.5\pm 1.9)D_{GC}+(7.6\pm 12.9)$ \citep{1994Wilson} is the abundance ratio of $\mathrm{^{12}CO}$ and $\mathrm{^{13}CO}$. For W49N, the Galactocentric distance $D_{GC}= 7.6$ kpc \citep{2020LiuTie}, therefore $\chi$ is approximately 64.6. By treating the opacity of $\mathrm{^{13}CO}$ as a function of velocity, we can perform channel-by-channel optical depth corrections. Here we assume a lower excitation temperature limit of 30 K \citep{2017Tang} to derive the lower bound of molecular mass. The momentum and kinetic energy can be derived by
\begin{equation}
    P = \sum M(v)v\times \frac{1}{\sin{i}}
\end{equation}
\begin{equation}
    E = \frac{1}{2}\sum M(v)v^2\times\frac{1}{\sin^2{i}},
\end{equation}
where $v$ is the LOS velocity relative to the systemic velocity, and $i$ is the inclination angle with respect to the plane of sky.

For the second approach utilizing $\mathrm{^{13}CO}$ emission, we can substitute specific parameters for $\mathrm{^{13}CO}$ into \autoref{eq:column_density} and \autoref{eq:gas_mass} to obtain the outflow mass, which follows
\begin{equation}
M_{13} = 1.08\times10^{-4}e^{\frac{15.87}{T}}(T+0.88)D^2\int{S_\nu dv}.
\end{equation}

The method for estimating momentum and kinetic energy is the same as in the previous case. The results (without inclination correction) are listed in \autoref{tab:OutflowParameters}, where the parameters obtained using $\mathrm{^{12}CO}$ and $\mathrm{^{13}CO}$ emission both assuming optically thin are labeled with the subscript ``12" and ``13" respectively, while those obtained through the optical depth correction method are labeled with ``12$\mathrm{\tau}$". Since in high-velocity channels the emission flux of $\mathrm{^{13}CO}$ is almost undetectable, and $\mathrm{^{12}CO}$ emission can be considered optically thin, no $\tau$-corrections were conducted in the highest velocity part shown in the first and last rows of the table. The estimated total outflow mass is over 130 $\mathrm{M_\odot}$, total momentum is over 3000 $\mathrm{M_\odot\ km\ s^{-1}}$, and outflow energy reaches the order of $10^{48}$ $\mathrm{erg}$, placing it close to the upper end of $10^{45-48}$ $\mathrm{erg}$ range reported for massive protostellar outflows in \cite{2002Beuther} and \cite{2015Maud}. Our results are of the same order of magnitude as those reported in \cite{1986Scoville} and \cite{2015LiuTie} for the same source. The momentum and energy of the molecular outflow reported here should be regarded as lower limits. This underestimation arises from four main factors: (1) the use of LOS velocities without inclination correction, (2) the adoption of a lower-bound excitation temperature of 30 K for the dense environment around W49N-G, (3) the incomplete spectral coverage of extremely high-velocity outflow, and (4) the missing flux inherent to interferometric observation. Even when considering the uncertainties of chemical abundance (0.3 dex for $X_{CO}$, see \cite{2013Bolatto}), conservative corrections suggest values exceeding our reported limits (maintaining the same order of magnitude): increasing $T$ to 50 K raises the parameters by $\sim33\%$, whereas adopting a mean inclination angle of 32.7$\arcdeg$ \citep[assuming random outflow orientations;][] {1996A&A...311..858B} would increase the intrinsic flow velocity by a factor of $\sim 2$ and the kinetic energy by a factor of $\sim 4$. This analysis provides robust evidence that this source drives one of the most energetic molecular outflows in the Galaxy, as suggested by \cite{2015LiuTie}.

\subsection{Kinematics and Dynamical Timescales of the Outflow Knots}\label{sec3.3}
\begin{figure*}[htbp]
\centering
\includegraphics[width=1\linewidth]{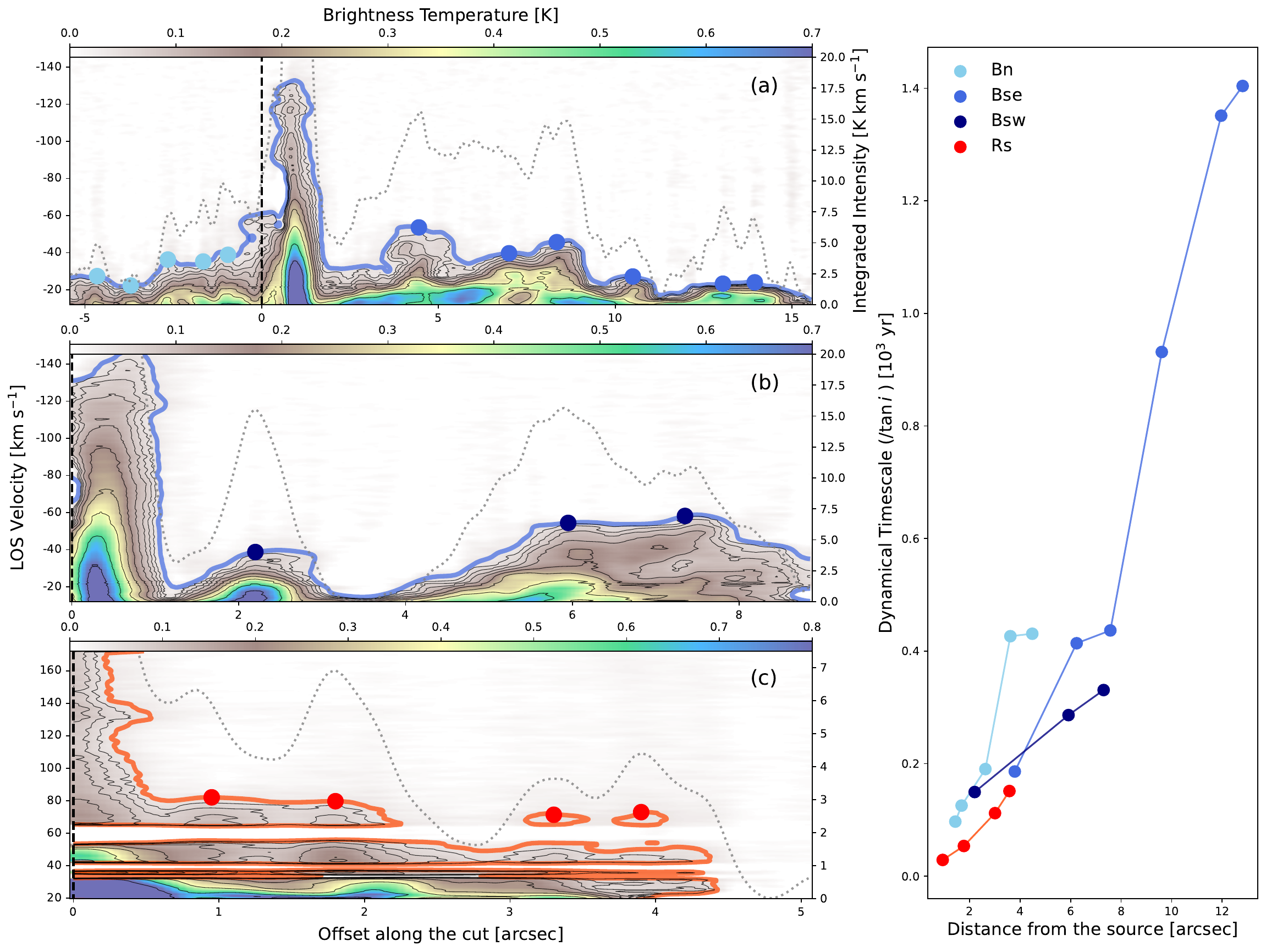}
\caption{\em{
\textbf{Left}: PV diagrams of $^{12}CO$ emission. The PV cuts of panels \textbf{(a)}, \textbf{(b)} and \textbf{(c)} are along the connection of the two curved arrows ``Bn" and ``Bse", the curved arrow ``Bsw" in \autoref{fig:channel_b} panel (a), and ``Rs" in \autoref{fig:channel_r} panel (a), respectively. Low velocity (less than \textasciitilde 10 $\mathrm{km\ s^{-1}}$ relative to the systemic velocity) parts are chopped out. The black dashed vertical lines indicate the position of source G2a. The smoothed contours in blue and red indicate 10 times the rms value of the PV diagram background. The colored dots on the smoothed contour denote the knots in four different newly found outflow lobes, which can be distinctly recognized by eye with the assistance of the grey-dotted integrated intensity curve. The intensity of the curve in panel \textbf{(c)} is integrated from the velocity channels higher than 40 $\mathrm{km\ s^{-1}}$ only to avoid contamination. \textbf{Right}: Dynamical timescale for each knot without inclination angle correction. The x-axis represents distance between each knot and the source.
}}
\label{fig:pvd}
\end{figure*}

To better investigate the kinematics of the observed outflow lobes, we extracted position-velocity (PV) diagrams along their propagation directions. Specifically, we chose PV cuts along the trajectories indicated by the arrows in \autoref{fig:channel_b} and \autoref{fig:channel_r}, focusing on the velocity ranges that show clear molecular outflows. Using the \textit{pvextractor} \footnote{https://pvextractor.readthedocs.io/en/latest/} Python package, we defined continuous, polygonal spatial paths tracking the central positions of the identified knots in the moment 0 maps. The PV slices were extracted directly along these specified trajectories, which allowed us to clearly trace the kinematic connectivity and emission structures between the discrete knots in position-velocity space. The resulting PV diagrams are presented in \autoref{fig:pvd}. Because the outflows exhibit complex structural characteristics across different channels, we visually identify the emission peak positions of the knots from the PV diagrams alongside their corresponding integrated intensity curves.

We then estimated the dynamical timescales ($t_{dyn}$) of the knots by utilizing their projected distances from the driving source and their LOS flow velocities. The LOS velocity associated with each knot was extracted directly from the PV diagrams. By tracing the prominent emission contours that outline the primary kinematic structure in the PV diagrams, we utilized linear interpolation to determine the precise velocity corresponding to the spatial position of each knot. The flow velocity for a given knot was then taken as the difference between this traced LOS velocity and the systemic velocity. Finally, the dynamical timescale was computed by dividing the projected physical distance of the knot by its flow velocity ($t_{dyn}/\tan{i}=d_{proj}/v_{flow}$). To correct for the inclination, the timescale estimate should be multiplied by $\tan{i}$.

The derived dynamical ages of the newly found outflow knots are mostly less than 1000 years (without inclination correction), suggesting that the central massive protostar is still in an early stage of formation. Furthermore, we estimated the timescale intervals between adjacent knots to characterize the periodicity of the episodic ejections. The estimated periods range from several decades to several centuries, which are similar to those obtained in low-mass star-forming regions \citep{2015Plunkett,2020A&ARv..28....1L,2024AJ....167...72D}.

\section{Discussion} \label{sec4}
Observational proof of jet-like lobes with knots in energetic multipolar molecular outflows like this one have not been commonly seen before. As derived in \autoref{sec3.3}, the dynamical ages of these newly identified outflow components suggest that the central massive protostar is still in an early stage of formation. 

\begin{figure*}
    \centering
    \includegraphics[width=0.8\linewidth]{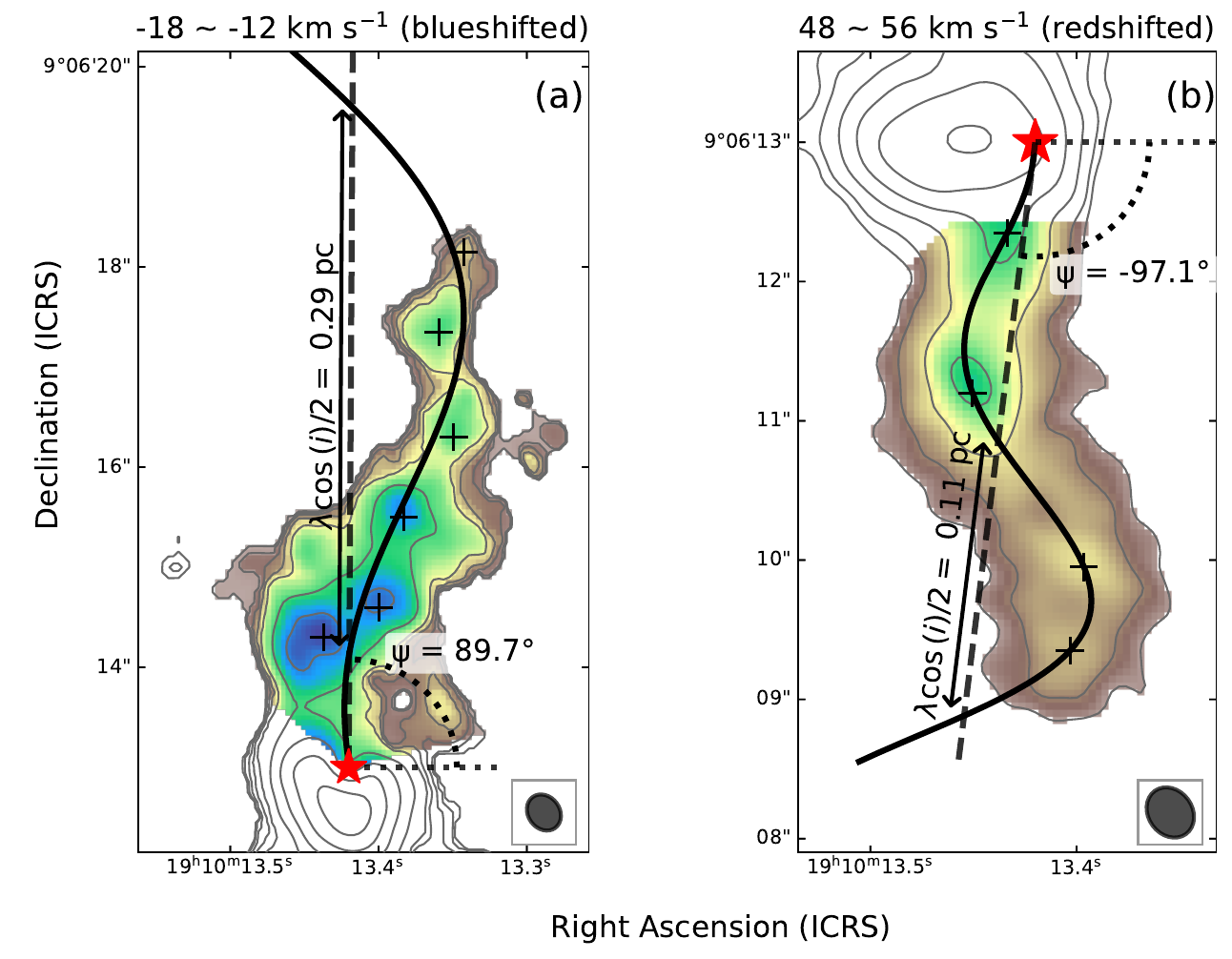}
    \caption{\em{Panel \textbf{(a):} Wiggle fitting of outflow lobe ``Bn". The color scale shows the integrated intensity within the region selected for fitting, overlaid on the full moment 0 contours. The region near the source was removed to avoid contamination from the compact outflow lobe with high inclination angle. The red star represents source G2a, consistent with that in \autoref{fig:channel_b}. The black solid line shows the outflow trajectory, and the dashed line represents the precession axis. The position angle of axis and projected half precession length scale are marked in the figure. Panel \textbf{(b):} Wiggle fitting of lobe ``Rs".}}
    \label{fig:wiggle}
\end{figure*}

As shown in \autoref{fig:channel_b} and \autoref{fig:channel_r}, knots are not arranged in a straight line in the blueshifted outflow lobe toward the north and the redshifted outflow lobe toward the south, but instead exhibit S-shaped wiggles (see the zoom-in view in \autoref{fig:wiggle}). This pattern may suggest the tidal effect in a system comprised of non-coplanar multiple protostars, which causes the precession of the accretion disk \cite[e.g.,][]{1999Terquem}. Since the inclination angles of lobes ``Bn" and ``Rs" are relatively small, and the sky projection does not severely deviate from a sinusoidal shape, we can use a two-dimensional approximate precession model of the form
\begin{equation}
    \begin{pmatrix}
    x  \\
    y 
    \end{pmatrix}
    =
    \begin{pmatrix}
    \cos{\psi} & -\sin{\psi} \\
    \sin{\psi} & \cos{\psi}
    \end{pmatrix}
    \times 
    \begin{pmatrix}
    \alpha l \sin{\left(\frac{2\pi l}{\lambda}+\phi_0\right)}  \\
    l \cos{i}
    \end{pmatrix}
\end{equation}
to fit the wiggles \citep{1996Eisloffel,2009Wu}. Here $\psi$ is the position angle of jet axis, measured counterclockwise from the west in the plane of the sky. $\alpha$, $\lambda$ and $\phi_0$ are amplitude, precession length scale and initial phase of outflow precession. $i$ is the inclination angle with respect to the plane of sky. $l$ is the distance from the source. Assuming these two lobes share the same intrinsic flow velocity of 160 $\mathrm{km\ s^{-1}}$ (derived from the high-velocity compact jet) for convenience, we estimate that the inclination angles for ``Bn" and ``Rs" are approximately 16$\arcdeg$ and 31$\arcdeg$, respectively. As shown in \autoref{fig:wiggle}, we selected the area with emission intensity greater than 3 times the rms noise (18.2 and 10.2 $\mathrm{Jy\ beam^{-1}\ km\ s^{-1}}$ for lobe ``Bn" and ``Rs", respectively) and cut the region contaminated by the central high-velocity compact outflow, and then located the emission peaks near the lobe centerline to perform geometric fitting. The fitted curves and directions of outflow axes are shown in \autoref{fig:wiggle}. From the fit, we derive inclination-corrected precession length scales of 0.61 pc and 0.25 pc for lobe ``Bn" and ``Rs", respectively. The corresponding amplitudes are $\sim$ 0.26 and $\sim$ 0.22, which imply angles of $\sim15\arcdeg$ and $\sim12\arcdeg$ between precession axis and outflow trajectory. Dividing the precession length scales by the projected velocities along the precession axes yields precession periods on the order of $\sim$ 1000 yr. This value is rough estimate, however, even with 50\% uncertainty in the inclination angles, the order of magnitude of the precession timescale remains robust.

\begin{figure*}[ht!]
    \centering
    \includegraphics[width=1\linewidth]
    {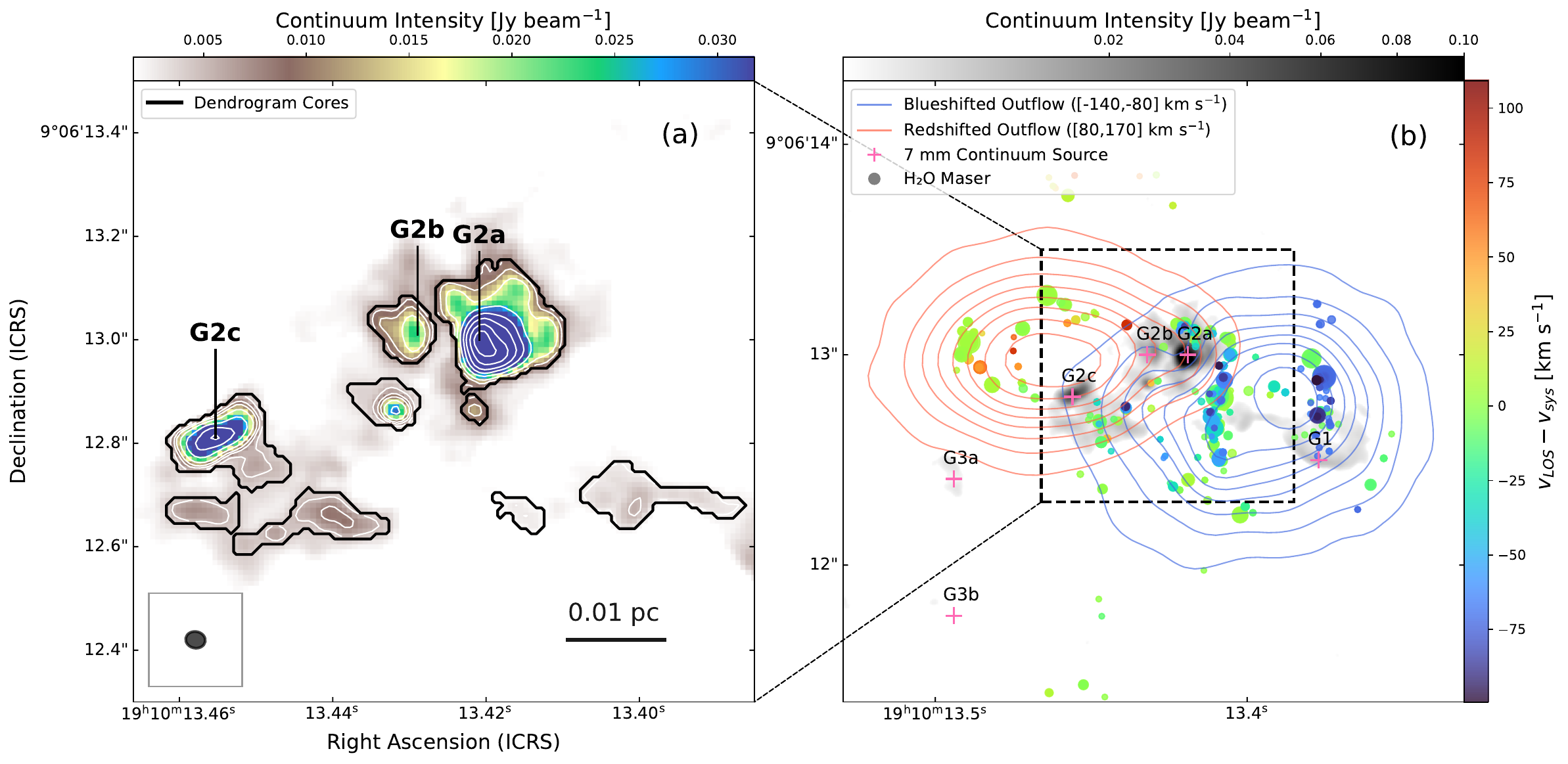}
    \caption{\em{Panel \textbf{(a):} A zoom-in view of the high-resolution 1.3 mm continuum emission from the ALMA archival data surrounding the sources G2a, G2b, and G2c, corresponding to the region marked by the dashed black box in the right panel. The thick black contours with labels outline the core boundaries identified by the astrodendro algorithm. The white contours represent the continuum emission levels at [5, 10, 15, 20, 30, 50, 80, 120, 180, 250] times the continuum rms noise ($5.3\times 10^{-4}\ \mathrm{Jy\ beam^{-1}}$). Panel \textbf{(b):} Water maser spotmap \citep[data from][]{2013ApJ...775...79Z} overlaid on moment 0 contours of extremely high-velocity bipolar outflow. The background is the high-resolution 1.3 mm continuum map on the center of W49N. The pink crosses represent 7 mm continuum objects observed in \cite{2020DePree}. Blue and red contours represent outflows with channel velocities of [-140,-80] and [80,170] $\mathrm{km\ s^{-1}}$ (consistent with the color maps in \autoref{fig:channel_b}(d) and \autoref{fig:channel_r}(d)), and the contour levels are [15,25,35,45,55,65,77,90,105] times the rms noises for redshifted and blueshifted moment-0 map, respectively. The spot sizes show the intensity of water maser emission and colors represent LOS relative velocity.}}
    \label{fig:maser}
\end{figure*}

In contrast to the ubiquitous bipolar outflows driven by low-mass protostars, the outflow from source G2 in W49N has up to six directions, but the outflow lobes are more jet-like without a very wide opening angle. This multipolar outflow could be driven by unresolved multiple protostars. To systematically map the continuum substructure within this protocluster and identify potential outflow driving sources, we extracted the cores using the \textit{astrodendro} \footnote{https://dendrograms.readthedocs.io/en/stable/} algorithm, a hierarchical decomposition of emission, on the high-resolution ($\sim0.03\arcsec$ or $\sim300$ AU) 1.3 mm continuum data (shown in \autoref{fig:maser}). We set the minimum pixel intensity threshold $min\_value=3\times rms_{cont}$, the minimum height for local peak as an independent structure $min\_delta=1\times rms_{cont}$, and the minimum number of pixels for the highest level of hierarchical structure $min\_npix=20$, where $rms_{cont}=5.3\times 10^{-4}\ \mathrm{Jy\ beam^{-1}}$ is the rms noise of continuum emission. The spatial positions of these extracted cores are in excellent agreement with the 7 mm continuum sources identified in \cite{2020DePree}. Morphologically, the convergence of the newly identified primary outflow lobes points most directly toward the vicinity of G2a and G2b. These two sources exhibit the strongest 1.3 mm continuum emission within the geometric origin region of the outflows, while dominated by free-free emission from ionized gas \citep{2020Nony}. Moreover, G2b seems a linear "jet" feature seen in 7 mm continuum, and thus may not be another prominent outflow driving source \citep{2000ApJ...540..308D}. Based on these characteristics, we conclude that the primary driving source of this outflow system is most likely a protostar (or multiple protostars) embedded within G2a. However, we cannot rule out the existence of other deeply embedded outflow driving sources within this protocluster, nor the possibility that the multipolar morphology is a superposition of independent monopolar or bipolar outflows from other continuum sources in this region. Future higher-resolution molecular line observations---particularly of high-excitation tracers---and sensitive centimeter-wave observations of radio jets are crucial to pinpoint the specific driving sources.

Under the current resolution of the archival continuum data, G2a is resolved (deconvolved size $\sim0.08\arcsec$ or $\sim900$ AU) but no binary is found within it. If a close binary system exists within G2a, the projected separation of the binary system should be smaller than the current resolution of 300 AU. Assuming a binary separation of 300 AU and a total mass of 30 $\mathrm{M_\odot}$ \citep[for an O8.5 star in G2a;][]{2020DePree}, we derived an orbital period of $\sim10^3$ yr. While a higher binary mass would slightly reduce this value ($P\propto M^{-0.5}$), the timescale remains robust and consistent with the previously derived outflow precession period, supporting the binary-driven scenario.

Another explanation for the multipolar outflow involves asymmetric accretion in the early phases of protostar formation within the extremely complex environment near W49N-G. Simulations by \cite{2020Rosen} suggest that the momentum and angular momentum fluctuations at early accretion can cause the protostar to move and induce rapid changes in the primary star's spin axis, leading to a precessing jet that manifests as a multipolar morphology. Alternatively, an explosive outflow event resulting from dynamical encounters within a high-density cluster presents another possible scenario \citep{2011Bally}. \cite{2009Smith} proposed Source G as a highly luminous analog to the OMC-1 explosive outflow. Such an explosive event could simultaneously eject cloudlets in multiple directions, naturally explaining the multipolar lobes. Due to the extreme distance to W49N and current resolution limits, however, a more detailed study on the ejecta and accretion process cannot be conducted to provide more solid conclusions. 

In \autoref{fig:maser} panel (b) we present the relationship between high velocity outflows and water maser spots, where maser spots are spatially surrounding the emission peak of the molecular outflow lobes. Given that the compact bipolar outflow directions are almost along the line of sight, this provides us with a unique, nearly ``tomographic" view, revealing the internal nested structures of the outflow and water masers. This structure is consistent with the ``jet-cocoon" model: a primary jet drives an expanding, high-pressure cocoon, and water masers are excited in the outer shock shell formed as the expanding cocoon pushes outward and interacts with the surrounding dense medium \citep{1994MacLow}. 

W49N-G2 is far from being quiescent. The continuum intensity has undergone a significant drop alongside a broadening of radio recombination lines \citep{2020DePree}, and the water masers exhibit intense flares \citep{2023Asanok}. These phenomena suggest episodic ejection driven by the central massive protostar(s). We speculate that the recent accretion burst of the central engine has ejected a new, high-speed stream of cloudlets. These ejecta subsequently chase and collide with the inner wall of the previously formed, expanding cocoon layer, and drive new strong shocks on the cavity wall of the cocoon, thereby triggering the outburst of the water masers.

To conclude, even though the jet-like outflows in W49N are extremely energetic, their properties (knotty structures, jet precession) are similar to those of low-mass protostars. This indicates that the underlying physical mechanisms for launching episodic outflows could be fundamentally similar in the formation of protostars across a wide range of stellar masses.

\section{Summary} \label{sec5}
The investigation of the molecular outflow in W49N, based on high-resolution $\mathrm{^{12}CO}$ (2--1) and $\mathrm{^{13}CO}$ (2--1) observations from the ALMA-QUARKS survey, has led to the following main conclusions:

1. The outflow is complex and multipolar, including four newly identified jet-like lobes. More reliable estimates of outflow parameters based on $\mathrm{^{12}CO}$ and $\mathrm{^{13}CO}$ emission (total mass \textgreater 130 $\mathrm{M_\odot}$, momentum \textgreater 3000 $\mathrm{M_\odot\ km\ s^{-1}}$, energy $\sim 10^{48}\ \mathrm{erg}$) confirm it to be one of the most energetic outflows in the Galaxy.

2. The new lobes exhibit distinct chains of knots, providing compelling evidence of episodic ejections with periods of decades to centuries, timescales similar to those observed in low-mass protostars. Two lobes (``Bn" and ``Rs") show prominent S-shaped wiggles, suggesting jet precession. Geometric fitting yields a precession period of $\sim1000$ yr, comparable to the orbital period of a potential binary system with a separation less than 300 AU within the G2a source.

3. The water maser spots are distributed surrounding the emission peaks of the high-velocity outflow lobes. This structure supports a ``jet-cocoon" model. We speculate that the observed intense maser flares are triggered by episodic ejections that drive strong shocks into the inner cavity wall of the previously formed cocoon.

4. The presence of episodic outflow and jet precession---features typical of low-mass protostars---in such an extreme high-mass environment points to a common underlying mechanism for outflow launching that spans a wide range of stellar masses.

\section{Acknowledgments}
T.L. acknowledges the supports by the National Key R\&D Program of China no. 2022YFA1603100, National Science and Technology Major Project 2024ZD1100601, National Natural Science Foundation of
China (NSFC) through grants no. 12073061 and no. 12122307, and the Tianchi Talent Program of Xinjiang Uygur Autonomous Region. W.J. acknowledges support from the Shanghai Post-doctoral Excellence Program. QY-L acknowledges the support by JSPS KAKENHI Grant Number JP23K20035. This research was carried out in part at the Jet Propulsion Laboratory, which is operated by the California Institute of Technology under a contract with the National Aeronautics and Space Administration (80NM0018D0004). AH thanks the support by the S. N. Bose National Centre for Basic Sciences under the Department of Science and Technology, Govt. of India and the CSIR-HRDG, Govt. of India for the funding of the fellowship. SRD acknowledges support from the Fondecyt Postdoctoral fellowship (project code 3220162) and ANID BASAL project FB210003. GG and LB gratefully acknowledge support by the ANID BASAL project FB210003.

This paper makes use of the following ALMA data: ADS/JAO.ALMA\#2021.1.00095.S and 2018.1.00520.S. ALMA is a partnership of ESO (representing its member states), NSF (USA) and NINS (Japan), together with NRC (Canada), NSTC and ASIAA (Taiwan), and KASI (Republic of Korea), in cooperation with the Republic of Chile. The Joint ALMA Observatory is operated by ESO, AUI/NRAO and NAOJ.

%

\facilities{ALMA, VLBA}
\software{Astropy \citep{astropy:2013,astropy:2018,astropy:2022}, Matplotlib \citep{matplotlib}, spectral-cube \citep{spectralcube}, pvextractor \citep{pvextractor}, astrodendro \citep{2008Rosolowsky}}





\bibliography{sample631}{}
\bibliographystyle{aasjournal}



\end{document}